\documentclass[prb,aps,twocolumn,superscriptaddress]{revtex4}
\usepackage{graphicx}
\usepackage{amsmath}
\usepackage{amssymb}
\usepackage{bm}
\newcommand{\rl}{\rangle\!\langle}
\newcommand{\tq}{\tilde{q}}
\DeclareMathOperator{\im}{Im}
\DeclareMathOperator{\re}{Re}
\begin{document}
\title{Theory of \textit{which path} dephasing in single electron interference 
due to trace in conductive environment}
\author{Pawe{\l} Machnikowski}
\email{Pawel.Machnikowski@pwr.wroc.pl} 
\affiliation{Institute of
Physics, Wroc{\l}aw University of Technology, 50-370 Wroc{\l}aw,
Poland} 

\begin{abstract}
A single-electron two-path interference (Young) experiment is
considered theoretically. 
The decoherence of an electron wave packet due to the
\emph{which path} trace left in the conducting (metallic) 
plate placed under the electron trajectories is
calculated using the many-body quantum description of the
electron gas reservoir. 
\end{abstract}


\maketitle

\section{Introduction}

Understanding the interaction between a microscopic quantum system 
and its macroscopic environment is one of the most exciting challenges
of the modern quantum mechanics. Such an interaction is known to
perturb the phase relations between the components of quantum
superpositions which damages the system ability to show quantum
interference. This effect, known as decoherence or dephasing, is
essential for our
understanding of the origins of classical
behavior.\cite{joos03,zurek03}
Moreover, it limits the feasibility of exploiting the
quantum properties of micro- and mesoscopic systems for performing
useful tasks, out of which quantum computation \cite{nielsen00,alber01}
is perhaps the widest known. 

Dephasing phenomena manifest themselves in any quantum system. It is,
however, natural to seek for their realization in the most clear
form. Such an ultimate demonstration seems to be a realization of 
a ``thought'' experiment opening many quantum mechanics
textbooks \cite{feynman3}: vanishing of the single-particle interference 
fringes in a two-slit (Young) experiment
due either to some perturbation of the environment state or to a
controlled measurement. Such a trace
(often referred to as \textit{which path} information) left in the
measurement device or in the
environment allows one (at least in principle) 
to identify the path chosen by 
the particle and destructs the interference to the extent depending 
on the distinguishability of the paths. 

Experiments of this kind were realized with electrons in a 
semiconductor system \cite{aleiner97,buks98a,avinun04} 
as well as in optical,\cite{zou91,brune96} neutron,\cite{summhammer87}
and atomic \cite{scully91,pfau94,chapman95,durr98}
interference setups. It was also suggested \cite{machnikowski05b}
that an analogous demonstration in the time domain could be based on 
interference experiments \cite{bonadeo98,kamada01,htoon02} 
in semiconductor quantum dots.

Another realization has been proposed,\cite{anglin96,anglin97} involving
free electron interference with a plate of finite conductivity
placed below the electron paths and parallel to them. 
A model for the decoherence effect in such a setup
has been formulated on phenomenological basis,\cite{anglin96} using the picture of
image charges screening the external flying electron. The image charge
would move parallel to the real electron inducing energy dissipation
(Joule heat generation) due to finite conductivity of the material.
The dissipation effect was quantitatively estimated
using an earlier calculation of velocity fields penetrating a 
metal.\cite{boyer74} The essential qualitative features of the resulting
dephasing effect is the dependence on the distance to the plate (which
governs the overall intensity of the \textit{which path} trace) 
and on the distance between
the paths (on which the degree of distinguishability depends).

With the state-of-the-art experimental techniques, such a free-electron
interference and controlled dephasing experiment is feasible. 
It is possible to observe interference fringes in experiments with
single electrons.\cite{merli76,tonomura89,hasselbach93}
It has been proposed to use a single-electron interferometer
\cite{hasselbach88} to show the controlled dephasing effect due to the 
conducting plate.\cite{hasselbach00,sonnentag05}

The present paper aims at the development of a fully quantum
description of the electron dephasing under the specific experimental
conditions as described above. 
On the microscopic level, the \emph{which path} information is
transferred from the electron (the system) to the conducting plate (the
environment, or reservoir) by exciting the electrons in the plate. 
This effect may be described by analyzing the electron coupling to the
electromagnetic modes affected by the presence of a conducting
surface, as previously done for a superconducting \cite{ford93} and
conducting \cite{levinson04} plate. In order to capture the essential
(dissipative) effect of the excitations in the reservoir it is,
however, profitable to formulate the description in terms of the
direct coupling between these two systems which is achieved by using
the Coulomb gauge.

A description 
developed in a different context \cite{alicki04a,grodecka05a} shows that
such a \textit{which path} decoherence effect can be quantitatively
represented as the overlap between the spectral density of the
reservoir fluctuations and the appropriate spectral function
related to the unperturbed evolution of the system. The present
calculation follows the same path, with the reservoir fluctuations
expressed in terms of the standard longitudinal dielectric function of
the conductor. It turns out that the interaction between the flying
electron and the
charges in the plate involves low frequencies but high momentum
transfer, which suggests that description going beyond the Debye model
is necessary. Therefore, the present calculation includes a complete 
model of the quantum properties of the reservoir. 

The paper is organized as follows. In the following Section
\ref{sec:model} the system under discussion and its model are
presented. 
Next, in Sec.~\ref{sec:general} the general framework of the theory is
described. In Sec.~\ref{sec:R} the problem is reduced to finding the
low frequency longitudinal dielectric function. This is done for
a metallic plate in the following Sec.~\ref{sec:metal}, where the 
quantitative results are found. The final Sec. \ref{sec:concl}
concludes the paper.

\section{The model}
\label{sec:model}

\begin{figure}[tb]
\begin{center}
\unitlength 1mm
\begin{picture}(65,30)(0,5)
\put(0,0){\resizebox{65mm}{!}{\includegraphics{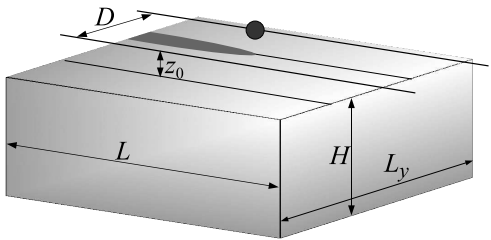}}}
\end{picture}
\end{center}
\caption{\label{fig:picture}Schematic plot of the system. The
information about the path chosen by the electron is transferred to
the plate by exciting the electron gas (local heating).}
\end{figure}

In the proposed experimental setup,\cite{hasselbach00,sonnentag05}
non-relativistic electrons with energies in the range of 150 eV to 3 keV are
emitted by an electron gun. 
This corresponds to velocities $v\sim 7\cdot 10^{6}$ to $3\cdot 10^{7}$
m/s (i.e., roughly one order of magnitude higher than the Fermi
velocities for metals \cite{kittel66}). 
The electron beam is split into two paths
separated by $D\sim 10\div 300$ $\mu$m. Both paths pass over a
conducting plate of length $L\sim 1$ cm
at a height $z_{0}\sim 0.1$ mm and then interfere on
a screen. The part of the experimental system relevant for the
dephasing effect is shown in Fig.~\ref{fig:picture}. 
It will be assumed that over the plate the electron paths are parallel.
The plate is taken to have a length $L$ and a very large height 
$H$ and width $L_{y}$. The
coordinate system is oriented in such a way that the $z=0$ plane 
coincides with the
plate surface and the electron paths are parallel to the $x$ axis.
In order to refer the present calculations directly to the experiment,
a single electron wave function will be chosen in the form of a
Gaussian quantum wave packet
travelling over the plate at a speed $v$ 
in a superposition of the two paths. 
We will assume that the electron wave packets 
corresponding to the two alternative paths do not overlap and that 
their extension is much
smaller than any distance in the experiment geometry. 
For the electron velocities as given above, the dispersion of a
minimum-uncertainty electron wave packet with initial width of a few
micrometers leads to an additional spread of $\sim 0.01\;\mu$m per
1~cm of flight path, so the effects of dispersion may be neglected.
It will also be assumed
that the time interval between the consecutive electrons is
much larger than the relaxation times relevant to the conducting plate
so that each flight event is independent of the previous ones. The
experiment is done at room temperature.

As mentioned above, contrary to the previous
work,\cite{ford93,levinson04}
the present analysis uses the Coulomb gauge. In
this way, the longitudinal electromagnetic field is eliminated in
favor of direct and instantaneous coupling to the electronic
degrees of freedom in the metal. Thus, the problem is of longitudinal 
screening type and may be treated using the well-established knowledge on
the density-fluctuation spectra of the electron gas and its
relation to the longitudinal dielectric function.\cite{note}
This choice relates also the present calculation more closely to
the original phenomenological concept \cite{anglin96} (although
the decoherence mechanism is described in a different
way).

The Hamiltonian for the Coulomb interaction of the
electron in the beam with the electrons and ions in the solid is
\begin{equation}
    V  =  \frac{e^{2}}{4\pi\epsilon_{0}}
      \int d^{3}r\int d^{3}r'
      \frac{\rho_{\mathrm{B}}(\bm{r})\rho(\bm{r}')}%
      {|\bm{r}-\bm{r}'|},\label{ham-full}
\end{equation}
where $\rho_{\mathrm{B}}(\bm{r})$
is the density operator for the electron beam,
$\rho(\bm{r})$ is the total charge density at a point $\bm{r}$ in the
conducting plate (including both electrons and ions), 
$e$ is the elementary charge, $\epsilon_{0}$ is the vacuum dielectric
constant (the SI system of units is used).
The ordering of operators in Eq.~(\ref{ham-full}) 
results from the assumption that the
electron above the plate is distinguishable from those in the
solid. Since a uniformly moving electron does not radiate transverse
fields on macroscopic distances, we disregard the coupling to the
electromagnetic vacuum via vector potential and calculate the contribution to
decoherence only from the longitudinal
excitations, thus neglecting the effect of the conducting surface via
modification of the transverse electromagnetic field 
fluctuations.\cite{levinson04,ford93} It should be noticed that we aim at a full
microscopic description of the interaction, hence no image potentials
are inserted ``by hand'' in the Hamiltonian (\ref{ham-full}). The
interaction between the electron and the charges in the plate
(classically described as the image potential) will
result from the microscopic theory, while the mutual screening
of electrons in the plate must be included in the correct description of the
longitudinal response of the interacting electron system.

Since we consider a flight of exactly one electron over 
the plate, the corresponding density operator may be written in the form
$\rho_{\mathrm{B}}(\bm{r})=|\bm{r}\rl\bm{r}|$, where $|\bm{r}\rangle$
is the position eigenstate of the electron. In the interaction
picture, the Coulomb part of the interaction Hamiltonian reads
\begin{equation}\label{ham-coul}
    V_{\mathrm{C}}(t)=\int d^{3}r
    U_{0}^{\dag}(t)|\bm{r}\rl\bm{r}|U_{0}(t)
    \frac{e^{2}}{4\pi\epsilon_{0}}\int d^{3}r'
      \frac{\rho(\bm{r}',t)}{|\bm{r}-\bm{r}'|},
\end{equation}
where $\rho(\bm{r},t)=U_{0}^{\dag}(t)\rho(\bm{r})U_{0}(t)$ and
$U_{0}(t)$ is the evolution operator for noninteracting subsystems.

The electron may travel along one of two paths,
with the corresponding quantum states
\begin{equation}\label{posrep}
    U_{0}(t)|0(1)\rangle=\int
    d^{3}r\psi_{\mathrm{0}(1)}(\bm{r},t)|\bm{r}\rangle,
\end{equation}
where $\psi_{\mathrm{0,1}}(\bm{r},t)$ are the electron wave functions in
the position representation.
Since the essential decoherence effect is related to path
distinguishability which affects only the relative phase between
these two quantum states, processes leading outside the
two-dimensional subspace spanned by these states may be neglected.
Projecting the electron states onto this subspace one gets
\begin{eqnarray*}
  \lefteqn{\left\langle 0(1)\left|
    V_{\mathrm{C}}(t)\right|0(1)\right\rangle =} \\
 && \frac{e^{2}}{4\pi\epsilon_{0}}
  \int d^{3}r |\psi_{\mathrm{0(1)}}(\bm{r},t)|^{2}
    \int d^{3}r' \frac{\rho(\bm{r}',t)}{|\bm{r}-\bm{r}'|},\\
  \lefteqn{\left\langle 1(0)\left|
    V_{\mathrm{C}}(t)\right|0(1)\right\rangle =} \\
 && \frac{e^{2}}{4\pi\epsilon_{0}}
  \int d^{3}r \psi^{*}_{\mathrm{1(0)}}(\bm{r},t)
        \psi_{\mathrm{0(1)}}(\bm{r},t)
    \int d^{3}r' \frac{\rho(\bm{r}',t)}{|\bm{r}-\bm{r}'|}=0,\\
\end{eqnarray*}
under the assumption that the wave functions corresponding to the
two basis states do not overlap. Hence, we have
\begin{eqnarray}\label{ham-2st}
    V_{\mathrm{C}}(t) & = & |0\rl 0|\frac{e^{2}}{4\pi\epsilon_{0}}
  \int d^{3}r \int d^{3}r' |\psi_{\mathrm{0}}(\bm{r},t)|^{2}
    \frac{\rho(\bm{r}',t)}{|\bm{r}-\bm{r}'|} \\
\nonumber
 && +|1\rl 1|\frac{e^{2}}{4\pi\epsilon_{0}}
  \int d^{3}r \int d^{3}r' |\psi_{\mathrm{1}}(\bm{r},t)|^{2}
    \frac{\rho(\bm{r}',t)}{|\bm{r}-\bm{r}'|}.
\end{eqnarray}

\section{The which path dephasing of the electron beam}
\label{sec:general}

In this Section the general framework for the description of the
perturbation to the electron state (in the 2-dimensional subspace
defined above) is described. The approach is based on the second order
expansion of the evolution equation for the density matrix in the
presence of the instantaneous Coulomb interaction with the charges in
the conducting plate. As a result one gets an expression for the
visibility of interference fringes in a single-electron interference
experiment in terms of a set of correlation constants depending on the
electron evolution and on the reservoir properties. Qualitative
discussion based on the general form of these correlation constants is
also presented here, while the quantitative analysis is performed in
the following Sections.

According to the standard theory,\cite{cohen98} 
the final reduced density matrix of the electron subsystem may be
written as
\begin{displaymath}
    \varrho(t)=U_{0}(t)\tilde{\varrho}(t)U_{0}^{\dag}(t),
\end{displaymath}
where $\tilde{\rho}(t)$ is the reduced density matrix in the
interaction picture. In the second order approximation (valid
as long as the overall perturbation effect is weak) the latter reads
\begin{eqnarray}\label{perturb}
    \tilde{\varrho}(t) & = & \varrho_{0}+\frac{i}{\hbar}
	\int_{t_{0}}^{t} d\tau
\mathrm{Tr}_{\mathrm{R}}[V_{\mathrm{C}}(\tau),
\varrho_{0}\otimes\varrho_{\mathrm{R}}]\\
\nonumber
 && 
-\frac{1}{\hbar^{2}}\int_{t_{0}}^{t} d\tau \int_{t_{0}}^{\tau} d\tau'
    \mathrm{Tr}_{\mathrm{R}}
    [V_{\mathrm{C}}(\tau),[V_{\mathrm{C}}(\tau'),
\varrho_{0}\otimes\varrho_{\mathrm{R}}]].
\end{eqnarray}
Here $\mathrm{Tr}_{\mathrm{R}}$ denotes the trace over the reservoir
degrees of freedom (i.e., electrons in the plate) and 
it has been assumed that at the initial time the system state is
separable into the product
$\varrho_{0}\otimes\varrho_{\mathrm{R}}$, where $\varrho_{0}$ is the
initial state of the electron and
$\varrho_{\mathrm{R}}$ is the thermal equilibrium state of the
electron gas in the conducting plate (this results from the long time
interval assumption).

Since the average fluctuation of the electron gas density in the
plate vanishes in equilibrium, the leading order contribution to the dephasing is
the second term. It may be written in the form
\begin{equation}\label{Dro}
    \Delta\tilde{\varrho}=
    -i[h_{t},\varrho_{0}]-\frac{1}{2}\{A,\varrho_{0}\}
	+\hat{\Phi}[\varrho_{0}],
\end{equation}
where the first contribution is a unitary correction and the two
other ones describe dephasing ($\{\cdot,\cdot\}$ denotes an
anti-commutator). In the present case it may be shown
that $h_{t}\sim\mathbb{I}$, so that this term does not contribute.
Physically, it contains the correction to the electron motion due to
the interaction with the image charge induced in the metal. Due to the
symmetry between the paths, these corrections are identical for both
states and do not induce any nontrivial evolution within the
restricted subspace. Their overall effect may be estimated by
considering the attraction force $e^{2}/(16\pi\epsilon_{0}z_{0}^{2})$ 
acting during the time of $L/v$, which results in a negligible vertical shift of
at most 100 nanometers.

The operator $A$ and the superoperator $\hat{\Phi}$ are
\begin{displaymath}
    A=\sum_{i=0,1}R_{ii}|i\rl i|,\;\;\;
    \hat{\Phi}[\varrho]=\sum_{ij=0,1}R_{ji}|i\rl i|\varrho|j\rl j|,
\end{displaymath}
where the correlation constants are
\begin{eqnarray}\label{Rij}
    R_{ij} & = & \frac{e^{4}}{(4\pi\epsilon_{0}\hbar)^{2}} \\
\nonumber
 && \times  \int_{t_{0}}^{t}d\tau\int_{t_{0}}^{t}d\tau'
   \int d^{3}r_{1}\int d^{3}r'_{1}\int d^{3}r_{2}\int d^{3}r'_{2} \\
\nonumber
 && \times
  \frac{|\psi_{i}(\bm{r}_{1},\tau)|^{2}}{|\bm{r}_{1}-\bm{r}'_{1}|}
    \frac{|\psi_{j}(\bm{r}_{2},\tau')|^{2}}{|\bm{r}_{2}-\bm{r}'_{2}|}
    \langle \rho(\bm{r}'_{1},\tau-\tau')\rho(\bm{r}'_{2})\rangle.
\end{eqnarray}
Note that the integration in Eq.~(\ref{Rij}) is symmetric in the time
variables, while the antisymmetric part has been separated into
$h_{t}$ in Eq.~(\ref{Dro}) (see Ref.~\onlinecite{grodecka05a} for technical
details). Obviously, $R_{ij}=R_{ji}^{*}$; in fact, both these constant
are real and equal to each other due to the symmetry between the beams
which excludes any relative phase shifts.

Although due to the finite size of the conducting plate $\langle
\rho(\bm{r}'_{1},\tau)\rho(\bm{r}'_{2})\rangle$ does not have the
full translational symmetry, in the transverse ($y$) direction it
may depend only on $y_{2}-y_{1}$. Hence, if the wave functions
$\psi_{\mathrm{0,1}}(\bm{r})$ differ only by a shift along $y$
then $R_{\mathrm{00}}=R_{\mathrm{11}}$. The evolution conserves
the diagonal elements of the density matrix, while the off-diagonal
ones change according to
\begin{displaymath}
    \langle 0|\Delta\tilde{\varrho}| 1\rangle
        =\Delta R \langle 0|\varrho_{0}|1\rangle,\;\;\;\
        \Delta R=R_{01}-R_{11}.
\end{displaymath}

Neglecting the plate edge effects (homogeneous approximation: the
decoherence effects simply accumulates while the electron is flying
over the plate) and
in the absence of reservoir memory (Markovian limit) the system
state is separable at any time, so that the above formula holds
for any time step. Then one expects that $\Delta R$ should be
proportional to the path segment $\Delta l$ traveled by the
electron and the quantity $\lambda^{-1}=-\Delta R/\Delta l$ becomes the
dephasing rate per unit path length. In this case, starting from
the equal superposition state
$\varrho_{0}=|\psi_{0}\rangle\!\langle\psi_{0}|$,
$|\psi_{0}\rangle=(|0\rangle+|1\rangle)/\sqrt{2}$, one gets after
the flight over the plate of length $L$
\begin{displaymath}
    \tilde{\varrho}=\frac{1}{2}(|0\rangle\!\langle 0|+|1\rangle\!\langle 1|)
       +\frac{1}{2}e^{-L/\lambda}
            (|0\rangle\!\langle 1|+|1\rangle\!\langle 0|).
\end{displaymath}
The detection probability for an electron at point $\bm{r}$ is
proportional to
\begin{eqnarray*}
I(\bm{r},t) &=&\langle \bm{r}|\varrho(t)|\bm{r}\rangle
=\langle \bm{r}
|U_{0}(t)\tilde{\varrho}(t)U_{0}^{\dag}(t) 
|\bm{r}\rangle\\
& = &
\frac{1}{2}\left\{  |\psi_{0}(\bm{r},t)|^{2} 
+|\psi_{1}(\bm{r},t)|^{2} \phantom{e^{-L/\lambda}}\right. \\
&&\left. +e^{-L/\lambda}\left[ \psi_{0}^{*}(\bm{r},t)\psi_{1}(\bm{r},t)
+\mathrm{H.c.}\right]
\right\},
\end{eqnarray*}
where we used the position representation defined in Eq.~(\ref{posrep})
At a point where the intensity $j(\bm{r})$ of the two beams is equal, i.e., 
$\psi_{1,2}(\bm{r},t)=\sqrt{j(\bm{r})}e^{i\phi_{1,2}(\bm{r},t)}$, one
has 
\begin{displaymath}
I(\bm{r},t)
=2j(\bm{r})\left[ 1+ e^{-L/\lambda}\cos\Delta\phi(\bm{r}) \right],
\end{displaymath}
where $\Delta\phi(\bm{r})=\phi_{1}(\bm{r},t)-\phi_{2}(\bm{r},t)$.
In the two-slit experiment this phase difference does not depend on
time but varies from point to point as a result of the difference of
the corresponding paths lengths, which leads to the interference
picture (it is assumed that $j(\bm{r})$ varies slowly in space).
The visibility of the interference fringes, defined in the standard
way using the maximum and minimum values of $I(\bm{r})$, is then
\begin{displaymath}
    \alpha=\frac{I_{\mathrm{max}}-I_{\mathrm{min}}}%
        {I_{\mathrm{max}}+I_{\mathrm{min}}}=e^{-L/\lambda}.
\end{displaymath}

Before presenting the detailed calculations it may be worthwhile
to discuss the effect qualitatively. First, if the
correlations are strongly local,
\begin{eqnarray*}
\lefteqn{R(\bm{r}_{1},\bm{r}_{2},t) =}\\ 
&&        \frac{e^{4}}{(4\pi\epsilon_{0}\hbar)^{2}}
    \int d^{3}r_{1}'\int d^{3}r_{2}'
    \frac{\langle \rho(\bm{r}'_{1},t)\rho(\bm{r}'_{2})\rangle}%
    {|\bm{r}_{1}-\bm{r}'_{1}||\bm{r}_{2}-\bm{r}'_{2}|} \\
 & = & \delta(\bm{r}_{1}-\bm{r}_{2})\tilde{R}(\bm{r}_{1},t),
\end{eqnarray*}
then
\begin{eqnarray*}
    R_{ij} & = & \int_{t_{0}}^{t}d\tau\int_{t_{0}}^{\tau}d\tau'
    \int d^{3}r_{1} \\
 && \times |\psi_{i}(\bm{r}_{1},\tau)|^{2}|\psi_{j}(\bm{r}_{1},\tau)|^{2}
    \tilde{R}(\bm{r}_{1},\tau-\tau')\sim \delta_{ij},
\end{eqnarray*}
since different wave functions do not overlap. In this case,
$\Delta R$ does not vanish and a dephasing effect appears.

On the other hand, for infinitely long-range correlations, one can write
\begin{displaymath}
    R(\bm{r}_{1},\bm{r}_{2},t)=f(\bm{r}_{1})f(\bm{r}_{2})
    \tilde{R}(t),
\end{displaymath}
where $f(\bm{r})=f(x,z)$ describes the dependence of the
interaction on the electron position relative to the plate. Then
\begin{eqnarray*}
    R_{ij} & = & \int_{t_{0}}^{t}d\tau\int_{t_{0}}^{t}d\tau'\left[
    \int dxdz \Phi(x,z,\tau)f(x,z)\right] \\
 && \times R(\tau-\tau')
    \left[ \int dxdz \Phi(x,z,\tau')f(x,z)\right],
\end{eqnarray*}
where
\begin{displaymath}
    \Phi(x,z,\tau)=\int dy |\psi_{\mathrm{0}}(\bm{r},\tau)|^{2}
        =\int dy |\psi_{\mathrm{1}}(\bm{r},\tau)|^{2},
\end{displaymath}
since the wave functions differ only by a shift along $y$.
Therefore $R_{\mathrm{01}}=R_{11}$ and the dephasing effect
vanishes.

In order to find quantitatively the decoherence path $\lambda$ the
correlation constants $R_{ij}$ must be evaluated using the correct
description of the properties of the electron gas in the conducting
plate. This is the subject of the following Sections.

\section{The correlation constants for the electron gas screening response}
\label{sec:R}

The goal of the present Section is to express the correlation
constants defined in Eq.~(\ref{Rij}) in terms of a standard
material response function (dielectric function) and a spectral
function pertaining to the unperturbed evolution (free flight) of
a single electron over the conducting plate. Since the dielectric
function is expressed in the momentum space and frequency domain,
Eq.~(\ref{Rij}) must be first Fourier-transformed.

As discussed in Sec.~\ref{sec:model}, 
the electron states are described by Gaussian wave packets
moving with the velocity $v$, localized along two parallel paths,
separated by a distance $D$, at a fixed distance $z_{0}$ from the plate,
\begin{displaymath}
    \psi_{i}(\bm{r},t)=\frac{1}{\pi^{3/2}l_{x}l_{y}l_{z}}
    e^{-\frac{1}{2}\left[ \frac{(x-vt)^{2}}{l_{x}^{2}}
    +\frac{(y-y_{i})^{2}}{l_{y}^{2}}+\frac{(z-z_{0})^{2}}{l_{z}^{2}}
    \right]},
\end{displaymath}
for $i=\mathrm{0,1}$, where $y_{\mathrm{0,1}}=\pm D/2$ and
$l_{x,y,z}$ are the wave function widths in the three directions.
%Defining the spatial Fourier transform
%\begin{displaymath}
%    \psi_{i}(\bm{p},t)=\int d^{3}r |\psi_{i}(\bm{r},t)|^{2}
%    e^{i\bm{p}\cdot\bm{r}}
%\end{displaymath}
For such a Gaussian state
one has
\begin{widetext}
\begin{equation}\label{psi-r-r}
  \int d^{3}r |\psi_{i}(\bm{r},t)|^{2}
  \frac{1}{|\bm{r}-\bm{r}'|} =
    \frac{1}{2\pi^{2}}\int \frac{d^{3}p}{p^{2}} e^{-i\bm{p}\cdot\bm{r}'}
    e^{i(p_{x}vt+p_{y}y_{i}+p_{z}z_{0})}
    e^{-\frac{1}{4}\left[(l_{x}p_{x})^{2}+(l_{y}p_{y})^{2}
    +(l_{z}p_{z})^{2}  \right]}.
\end{equation}

The conducting plate occupies the volume defined by $-L/2<x<L/2$,
$-L_{y}/2<y<L_{y}/2$, $-H<z<0$, where $L$ is the finite plate
length while $L_{y}$ and $H$ are assumed to be very large. Using the
fluctuation-dissipation theorem \cite{breuer02,mahan00}, the
density-density correlation function may be expressed by the
imaginary part of the full dielectric function of the conducting plate
\begin{equation}\label{dens-cor}
   \langle \rho(\bm{r}'_{1},t)\rho(\bm{r}'_{2})\rangle =
    \frac{\hbar \epsilon_{0}}{\pi e^{2}}\frac{1}{V}\sum_{\bm{q}}
    q^{2}\int_{0}^{\infty}d\omega
    e^{i\bm{q}\cdot(\bm{r}'_{2}-\bm{r}'_{1})}e^{-i\omega t}
    \coth \frac{\hbar\beta\omega}{2}
    \im\left[-\frac{1}{\varepsilon(\bm{q},\omega)}\right],
\end{equation}
where the expansion in space is made in terms of the discrete set
of plane waves in the finite volume of the plate (with periodic
boundary conditions). 

Let us assume that the total dephasing accumulated during the whole
flight over the plate is weak. In this case, the perturbative
Eq.~(\ref{perturb}) yields a
good approximate description of the effect. Since the interaction
takes place only when the electron is over the plate, one may shift
the initial and final times of the evolution $t_{0}$ and $t$ in
Eqs.~(\ref{perturb}) and (\ref{Rij}) to $-\infty$ 
and $+\infty$, respectively. Then,
substituting Eqs.~(\ref{psi-r-r}) and (\ref{dens-cor}) into
Eq.~(\ref{Rij}) and performing the integral over times we
get
\begin{eqnarray*}
    R_{ij} & = & \frac{e^{2}}{16\pi^{5}\hbar\epsilon_{0}}\frac{1}{V}
    \sum_{\bm{q}}q^{2}\int_{0}^{\infty}d\omega
    \coth \frac{\hbar\beta\omega}{2}
    \im\left[-\frac{1}{\varepsilon(\bm{q},\omega)}\right]
    \int d^{3}r_{1}'\int d^{3}r_{2}' \int \frac{d^{3}p}{p^{2}}
     \int \frac{d^{3}p'}{{p'}^{2}}\\
 && \times
    \delta(p_{x}v-\omega)
    e^{i[-(\bm{p}+\bm{q})\cdot\bm{r}'_{1}+p_{y}y_{i}+p_{z}z_{0}]}
    e^{-\frac{1}{4}\left[(l_{x}p_{x})^{2}+(l_{y}p_{y})^{2}
    +(l_{z}p_{z})^{2}  \right]} \\
 && \times
    \delta(p'_{x}v-\omega)
    e^{i[-(\bm{p'}-\bm{q})\cdot\bm{r}'_{2}+p'_{y}y_{j}+p'_{z}z_{0}]}
    e^{-\frac{1}{4}\left[(l_{x}p'_{x})^{2}+(l_{y}p'_{y})^{2}
    +(l_{z}p'_{z})^{2}  \right]}.
\end{eqnarray*}
Since in the $y$ direction the plate is very long, the integrals
over $y_{1,2}'$ yield a Dirac $\delta$. On the other hand, in the $x$
direction, the plate length is limited. Performing the four
integrations over $x_{1,2}',y_{1,2}'$, 
followed by those over $p_{x,y}$ and $p'_{x,y}$ one
arrives at
\begin{eqnarray}\label{Rij2}
  R_{ij} &=& \frac{e^{2}}{4\pi^{3}\epsilon_{0}\hbar v^{2}}
    \frac{1}{V}\sum_{\bm{q}}q^{2}\int_{0}^{\infty}d\omega
    \coth \frac{\hbar\beta\omega}{2}
    \im\left[-\frac{1}{\varepsilon(\bm{q},\omega)}\right]  \\
&& \nonumber \times
    e^{iq_{y}(y_{i}-y_{j})} e^{-\frac{1}{2}\left[
    \left( \frac{l_{x}\omega}{v}\right)^{2}+(l_{y}q_{y})^{2} \right]}
    2\pi L \delta_{L}\left( q_{x}+\frac{\omega}{v}\right)
    |I_{z}|^{2}.
\end{eqnarray}
Here
 $\delta_{L}(q)=\frac{4\sin^{2}\frac{L}{2} q}{2\pi L q^{2}}$
and
\begin{eqnarray*}
    I_{z} & = & \left[
    \int dp_{z}\frac{1}{(\omega/v)^{2}+q_{y}^{2}+p_{z}^{2}}
    e^{-\frac{1}{4}(l_{z}p_{z})^{2}}
        \int_{-H}^{0}dz e^{-i(p_{z}+q_{z})z+ip_{z}z_{0}}
        \right]_{H\to\infty} \\
 & = & \frac{1}{iq_{z}-\tq }\frac{\pi}{2\tq }
    e^{\frac{1}{4}(\tq l_{z})^{2}-\tq z_{0}}
    \left[1- \mathrm{Erf}\left(
         \frac{\tq l_{z}}{2}-\frac{z_{0}}{l_{z}} \right)
    \right],
\end{eqnarray*}
\end{widetext}
where $\tq=[q_{y}^{2}+(\omega/v)^{2}]^{1/2}$. In this result, terms
proportional to $(l_{z}/z_{0})\exp[-(z_{0}/l_{z})^{2}]$ have
been neglected, since $l_{z}\ll z_{0}$.

Since the characteristic momentum scales of the system are at
least of the order of $1/z_{0}$ and $z_{0}\ll L$, the broadening of
the function $\delta_{L}(q)$, which is of the order of $1/L$, may be
neglected and one can write $\delta_{L}(q)\approx \delta(q)$. In this
way one neglects the corrections related to the
approach to the edge of the finite plate and to the fly-away phase after
crossing it, compared to the dephasing accumulated during the flight
directly over the plate. It should be noted that Eq.~(\ref{Rij2}) is
limited to weak dephasing but it does not involve any Markovian
approximations. On the other hand, the memory time of the screening
response of the electrons in metals is of order of inverse Fermi
energy, i.e., femtoseconds, which is many orders of magnitude shorter
than any time scale of the problem. Vanishing memory together with the
proportionality of dephasing to the length travelled over the plate
allow us to interpret the perturbative result as a dephasing rate (per
unit path length) and to obtain an exponential decay as a solution of
the corresponding rate equation. As discussed in the
Appendix~\ref{app:M}, the same result is obtained by
coarse-graining the flight over the plate and explicitly using the
short memory assumption, which directly leads to correlation constants
$R_{ij}$ proportional to time over each small time step.

\begin{figure}[tb]
\begin{center}
\unitlength 1mm
\begin{picture}(85,40)(0,10)
\put(0,0){\resizebox{85mm}{!}{\includegraphics{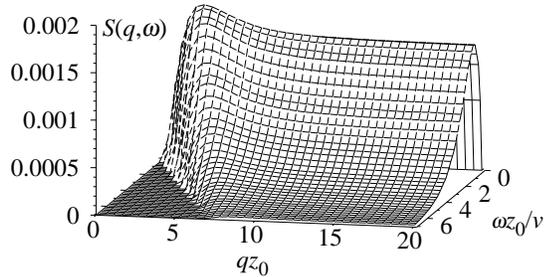}}}
\end{picture}
\end{center}
\caption{\label{fig:sqw}The form of the function $S(q,\omega)$ for
$D/z_{0}=0.1$ and $l_{x}=l_{y}=l_{z}=0.01z_{0}$.}
\end{figure}

Using Eq.~(\ref{Rij2}) in the above approximation and replacing the
summation over $\bm{q}$ by integration one may write
\begin{eqnarray}\label{DRij}
\Delta R & = & -\frac{e^{2}L}{\epsilon_{0}\hbar v}\frac{1}{(2\pi)^{3}}
\int\frac{d\omega}{\omega}\coth\frac{\hbar\beta\omega}{2} \\
\nonumber
&& \times \int \frac{d^{3}q}{q^{2}}\im\left[ 
 -\frac{1}{\varepsilon(\bm{q},\omega)} \right] S(\bm{q},\omega),
\end{eqnarray}
where 
\begin{widetext}
\begin{eqnarray*}
S(\bm{q},\omega) & = & \frac{\omega}{8vq}
\delta\left(\sin\theta+\frac{\omega}{qv}\right)
\left[ 1-e^{iDq\cos\varphi\cos\theta_{0}} \right] \\
&&
\times 
e^{-\frac{1}{2}\left( \frac{l_{x}\omega}{v} \right)^{2} 
   -\frac{1}{2}(l_{y}q)^{2}\cos^{2}\varphi\cos^{2}\theta_{0}
   -\left[ 2z_{0}q +\frac{1}{2}(ql_{z})^{2}\right]
	\sqrt{\cos^{2}\varphi\cos^{2}\theta_{0}
	+\left( \frac{\omega}{vq} \right)^{2} }}\\
&& \times\frac{1}{\cos^{2}\varphi\cos^{2}\theta_{0}
	+\left( \frac{\omega}{vq} \right)^{2}}
\left[ 1-\mathrm{Erf}\left( 
\frac{ql_{z}}{2}\sqrt{\cos^{2}\varphi\cos^{2}\theta_{0}
	+\frac{\omega^{2}}{v^{2}q^{2}} }-\frac{z_{0}}{l_{z}}
 \right)  \right]^{2},
\end{eqnarray*}
where we write 
$\bm{q}=q(\sin\theta,\cos\theta\cos\varphi,\cos\theta\sin\varphi)$ and
denote $\theta_{0}=\mathrm{arcsin}[\omega/(vq)]$. 
Since $z_{0}$ is a large (macroscopic) distance, for $q\gg
\omega/v\sim 1/z_{0}$
the above function is strongly peaked around $\varphi=\pm \pi/2$. 
On the other hand, the dielectric
function of a conductor extends over momenta of the order of the Fermi
momentum, $q\sim k_{\mathrm{F}}\gg 1/z_{0}$ and its imaginary part
vanishes for low momenta, so that for the relevant
momentum values one can write
\begin{equation}\label{S}
S(\bm{q},\omega)=\frac{1}{2}S(q,\omega)
\delta\left(\sin\theta+\frac{\omega}{qv}\right)
\left[ \delta\left( \varphi-\frac{\pi}{2} \right)
	 +\delta\left( \varphi+\frac{\pi}{2}\right) \right],
\end{equation}
where
\begin{displaymath}
S(q,\omega)=
\int_{0}^{2\pi}d\varphi\int_{-\pi/2}^{\pi/2}\cos\theta d\theta
S(\bm{q},\omega).
\end{displaymath}
Physically, the above formulas mean that the momentum transfer from
the flying electron to the electron gas excitations in the $(x,y)$
plane may be at most of order of $1/z_{0}$ and the energy transfer of
order of $\hbar v/z_{0}$. 

The integral over $\theta$ is trivial while that over $\varphi$
may be performed by a saddle point approximation around
$\varphi=\pm\pi/2$,
writing $\cos(\varphi\pm\pi/2)\approx\pm\varphi$ which, 
upon further substitution $\varphi=[(qv/\omega)^{2}-1]^{-1/2}u$ and
extending the integration limits, leads to
\begin{eqnarray*}
S(q,\omega)& = &\left[1-\left( \frac{\omega}{qv} \right)^{2} \right]^{-1/2}
\int_{-\infty}^{\infty}\left[ 1-e^{-i\frac{D\omega}{v}u} \right] 
\frac{1}{1+u^{2}}e^{-2\frac{z_{0}\omega}{v}\sqrt{1+u^{2}}}\\
&& \times
e^{-\frac{1}{2}\left( \frac{l_{x}\omega}{v} \right)^{2} 
   -\frac{1}{2}\left( \frac{l_{y}\omega}{v} \right)^{2}u^{2}
   -\frac{1}{2}\left( \frac{l_{z}\omega}{v} \right)^{2}\sqrt{1+u^{2}}}\frac{1}{4}\left[ 1-\mathrm{Erf}\left( 
\frac{l_{z}\omega}{2v}\sqrt{1+u^{2}}-\frac{z_{0}}{l_{z}} \right)
	 \right]^{2}.
\end{eqnarray*}
\end{widetext}
The analysis of the spectral function $S(q,\omega)$ 
(see Fig. \ref{fig:sqw}) shows that it
decays exponentially for $\omega\gtrsim v/z_{0}$ while for a fixed
$\omega$ it varies with $q$ on an interval $\sim 1/z_{0}$ to reach 
a plateau
for $q\gg \omega/v$, where it attains a constant value, dependent only
on $\omega$,
\begin{displaymath}
S(q,\omega)\stackrel{q\gg\omega/v}{\longrightarrow}S(\omega).
\end{displaymath}

The description simplifies considerably if one assumes that 
the size of the electron wave packet is much smaller
than the distance to the conducting plate, i.e., $l_{x,y,z}\ll
z_{0}$. Note that keeping the finite wave packet size leads only to
negligible quantitative corrections. In particular, it does not
introduce any momentum cut-off.
In the approximation $l_{x,y,z}\to 0$, the asymptotic value may be 
written in the form
\begin{equation}\label{S-asympt}
S(\omega)=
\int_{-\infty}^{\infty}du
\left[ 1-e^{-i\frac{D\omega}{v}u} \right] 
\frac{1}{1+u^{2}}e^{-2\frac{z_{0}\omega}{v}\sqrt{1+u^{2}}}.
\end{equation}

\begin{figure}[tb]
\begin{center}
\unitlength 1mm
\begin{picture}(45,30)(0,5)
\put(0,0){\resizebox{44mm}{!}{\includegraphics{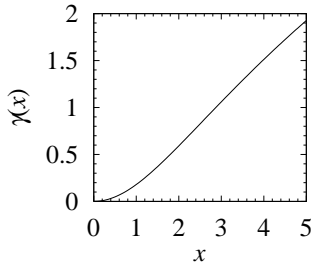}}}
\end{picture}
\end{center}
\caption{\label{fig:gamma}The ``geometrical'' function $\gamma(x)$.}
\end{figure}

In the low frequency range selected by the function $S(\omega)$, 
one has always
$\im\varepsilon^{-1}(\bm{q},\omega)\sim \omega$ (this is a general
property;\cite{mahan00} see also below).
Moreover, $\hbar v/(k_{\mathrm{B}}z_{0})\sim 1$ K, so that
at room temperature one may write 
$\coth(\hbar\beta\omega/2)\approx 2k_{\mathrm{B}}T/(\hbar\omega)$ for
all relevant frequencies $\omega\lesssim v/z_{0}$.
Therefore, the
frequency integral in Eq.~(\ref{DRij}) will always have the form
\begin{eqnarray}\label{gamma}
\int_{0}^{\infty}d\omega\frac{S(\omega)}{\omega}=
\gamma\left(\frac{D}{z_{0}}\right),
\end{eqnarray}
where
\begin{eqnarray}
\label{gamma-expli}
\gamma(x) & = & 
\frac{1}{2}\int_{-\infty}^{\infty}\frac{du}{1+u^{2}}\ln\left[ 
1+\frac{x^{2}}{4}\frac{u^{2}}{1+u^{2}} \right] \\
\nonumber
& = & \frac{\pi}{16} x^{2}+O(x^{4}),
\end{eqnarray}
[using the asymptotic form given in Eq.~(\ref{S-asympt})].
This function depends only on the geometry of the system. More
specifically, the only relevant parameter is the ratio $D/z_{0}$.
The function $\gamma(x)$ is plotted in Fig.~\ref{fig:gamma}. 

As for the momentum integral in Eq.~(\ref{DRij}), 
it obviously depends on the specific
properties of
excitations in the electron gas and may be very different depending,
for instance, on the relation between the thermal energy
$k_{\mathrm{B}}T$ and the Fermi energy, as well as on the conduction band
structure. In any case, however, this integral may be partly
performed using Eq.~(\ref{S}),
\begin{eqnarray*}
\lefteqn{\int\frac{d^{3}q}{q^{2}}
\im \left[ -\frac{1}{\varepsilon(\bm{q},\omega)} \right]}\\
&& 
\times 
\frac{1}{2}\delta\left(\sin\theta+\frac{\omega}{qv}\right)
\left[ \delta\left( \varphi-\frac{\pi}{2} \right)
	 +\delta\left( \varphi+\frac{\pi}{2}\right) \right] \\
& = & \int dq_{z} 
\im \left[ -\frac{1}{\varepsilon(\bm{q},\omega)} \right]_{q_{x}=q_{y}=0}, 
\end{eqnarray*}
where it is assumed again that $qv\gg\omega$. 
Hence, from Eq.~(\ref{DRij}), 
\begin{eqnarray}\label{DRfinal}
\lefteqn{\Delta R=}\\
\nonumber 
&& -\frac{k_{\mathrm{B}}Te^{2}L}{4\pi^{3}\epsilon_{0}\hbar^{2}v}
\gamma\left( \frac{D}{z_{0}} \right) \frac{1}{\omega}\int dq_{z}
\im\left[ -\frac{1}{\varepsilon(\bm{q},\omega)} 
\right]_{q_{x}=q_{y}=0}.
\end{eqnarray}
Since, at low frequencies,
$\epsilon_{2}(\bm{q},\omega)\ll\epsilon_{1}(\bm{q},\omega)$, one may
write
\begin{displaymath}
\im\left[ -\frac{1}{\varepsilon(\bm{q},\omega)} \right]\approx
\frac{\varepsilon_{2}(\bm{q},\omega)}{\varepsilon_{1}^{2}(\bm{q},\omega)}.
\end{displaymath}

Below, the results for metallic plates are presented.

\section{Decoherence over a metallic plate}
\label{sec:metal}

In this Section, the decoherence effect for a metallic
plate is considered. 
For the description of the reservoir response properties, we will
use the quasi-free electron (isotropic) longitudinal dielectric function in the
random phase approximation (the Lindhard theory).\cite{mahan00} 
By assuming periodic boundary conditions 
this dielectric function may be applied to a finite system with
negligible corrections as long as the typical momentum transfers are
much larger than the inverse system size, which is the case in our
problem. Obviously, the screening response originates only from the
finite volume of the plate, which has already been included in the
theory developed above. 

On the other hand, the Lindhard theory assumes that single-particle
states are just plain waves confined in a box with infinite potential
walls, which is not quite the case for a real piece of metal. In
particular, the dielectric discontinuity at the surface acts as a
trapping potential for carriers.\cite{pines89} Since the surface
states are confined in the vertical direction, low energy excitations
must involve in-plane momentum transfer, which is not favorable in our
problem. Thus, one should not expect large dissipation (dephasing) 
from these modes an their contribution will be dominated by virtual
high-frequency excitations (surface plasmons) which will
tend to screen the potential of the external charge, thus decreasing
the dephasing. This effect will be neglected in the present analysis. 

Even in this quasi-homogeneous approximation, 
the strictly correct description of the screening 
response requires solving the
complicated many-body problem, involving the electrons in the metal, 
the positive ions, and the mutual screening interaction between these
subsystems. Under the conditions of the present study
the treatment may be, however, greatly simplified with some loss of
precision but retaining the essential features and yielding reasonable
quantitative estimations. 

Thus, the calculation of the total dielectric function of a metal will be
done within the approximation \cite{mahan00} consisting in simply
adding the susceptibilities of the electron and ion subsystems, 
\begin{eqnarray*}
\epsilon(\bm{q},\omega) & = & 1+\chi_{\mathrm{el}}(\bm{q},\omega)
+\chi_{\mathrm{ph}}(\bm{q},\omega) \\
& = & [1+\chi_{\mathrm{ph}}(\bm{q},\omega)]
\left[1+\frac{1}{1+\chi_{\mathrm{ph}}(\bm{q},\omega)}
	\chi_{\mathrm{el}}(\bm{q},\omega)\right],
\end{eqnarray*}
where $\chi_{\mathrm{el}}(\bm{q},\omega),\chi_{\mathrm{ph}}(\bm{q},\omega)$
are the electron and phonon electric susceptibilities, respectively.

The phonon susceptibility, including the screening of the ion interaction
by electrons, is given by \cite{mahan00}
\begin{displaymath}
\chi_{\mathrm{ph}}(\bm{q},\omega)=
\frac{\omega_{\mathrm{pi}}^{2}}{\omega^{2}(\bm{q})-\omega^{2}},
\end{displaymath}
where $\omega_{\mathrm{pi}}=Ze[n/(\epsilon_{0}M)]^{1/2}$ is the ion
plasma frequency (here $Z$ is the ion charge, $n$ is the ion
concentration and $M$ is the ion mass). Since there is no momentum
cut-off in the problem under discussion, the system response is
dominated by excitations with $q$ of order of the Debye wave vector
$k_{\mathrm{D}}$, so that we may replace 
$\omega(\bm{q})$ by its short wavelength value for which the standard 
simple theory yields $\omega_{\mathrm{pi}}$. Also, since 
$\omega\sim v/z_{0}\ll\omega_{\mathrm{pi}}$, we may put $\omega\to 0$
in the denominator. In this approximation,
the effect of screening by the lattice excitations is fully contained 
in the ion dielectric constant
\begin{displaymath}
\epsilon_{\mathrm{i}}=1+\chi_{\mathrm{ph}}(k_{\mathrm{D}},0)\approx 2.
\end{displaymath}
In this discussion the imaginary part of the lattice
response describing the dissipative effects has been neglected. 
Such effects must
always be resonant, i.e., they involve phonons with low frequencies
$\omega\sim v/z_{0}$. The density of states for such long-wavelength
phonons is very low. Moreover, at the corresponding low momenta
$v/(c_{\mathrm{s}}z_{0})\ll k_{F}$, where $c_{\mathrm{s}}$ is the
sound speed, the real part of the total dielectric function is very
large due to electron contribution (see below), so that such
low-frequency lattice excitations are additionally very strongly
screened. Therefore, the lattice contribution to the dissipative
processes [i.e., to the $\varepsilon_{2}(\bm{q},\omega)$ function] is
negligible. In terms of the general picture of the dephasing due to
the trace in the environment this means that most of the phonons have
frequencies much higher than that characteristic of the external field
so that they follow the perturbation adiabatically, returning to the
original state after the flying electron is away and thus registering
no trace of its passage.

The total dielectric function including electron transitions within the
conduction band and the lattice screening as described above
is therefore described by the formula
\begin{eqnarray}\label{diel}
\varepsilon(\bm{q},\omega)=
\epsilon_{\mathrm{i}}\left[ 
1+\frac{1}{\epsilon_{\mathrm{i}}}
\chi_{\mathrm{el}}(q,\omega) \right] ,
\end{eqnarray}
where we take into account that in the isotropic model the dielectric 
function may depend only on the value but not on the direction of $\bm{q}$. 
The electron susceptibility in
the quasi-free electron model valid for arbitrary $q$ is given by the
Lindhard formula \cite{mahan00}
\begin{equation}\label{diel-expli}
\chi_{\mathrm{el}}(q,\omega)=
\frac{e^{2}}{4\pi^{3}\epsilon_{0}q^{2}}
\int d^{3}p 
\frac{n_{\mathrm{F}}(E_{\bm{p}})-n_{\mathrm{F}}(E_{\bm{p}+\bm{q}})}{
{E_{\bm{p}}}-{E_{\bm{p}+\bm{q}}}+\omega+i0^{+}},
\end{equation}
where $E_{\bm{p}}=p^{2}/(2m)$, $m$ is the electron mass and
$n_{\mathrm{F}}(E)$ is the Fermi-Dirac distribution.
Since the function $S(\bm{q},\omega)$ has an exponential
spectral cut-off at frequencies much lower than the Fermi energy, one
needs only the low-frequency part of the dielectric function (as
already remarked in Sec.~\ref{sec:R}).
To be specific, $\hbar v/z_{0}\sim 10^{-5}E_{\mathrm{F}}$, where
$E_{\mathrm{F}}$ is the Fermi energy (the relevant frequency range
corresponds to radio frequencies). Thus, for the imaginary part of
the susceptibility [Eq.~(\ref{diel-expli})] we write, 
using the standard results,\cite{mahan00}
\begin{equation}\label{Ime}
\im\chi_{\mathrm{el}}(q,\omega)
=  \left\{
\begin{array}{ll}
\frac{e^{2}m^{2}}{2\pi\epsilon_{0}\hbar^{3}q^{3}}\omega\; & 
	\mbox{for}\ \omega/v_{\mathrm{F}}<q<2k_{\mathrm{F}}, \\
0 & \mbox{otherwise}.
\end{array}
\right.
\end{equation}
The electron velocities used in the experiment satisfy $v\gg
v_{\mathrm{F}}$ for practically all metals.
Thus, the condition $q\gg\omega/v$ is automatically satisfied and the
asymptotic form of the spectral function given by Eq.~(\ref{S-asympt})
may always be used.

The leading low-frequency term of
the real part of Eq.~(\ref{diel-expli}) is
\begin{equation}\label{Ree}
\re\chi_{\mathrm{el}}(q,0)
=
 \frac{me^{2}k_{\mathrm{F}}}{
	2\pi^{2}\epsilon_{0}\hbar^{2}q^{2}}
\left(
1+\frac{4k_{\mathrm{F}}^{2}-q^{2}}{4qk_{\mathrm{F}}}
\ln \left| \frac{q+2k_{\mathrm{F}}}%
		{q-2k_{\mathrm{F}}} \right|
 \right),
\end{equation}
where $k_{\mathrm{F}}$ is the Fermi wave vector.

\begin{figure}[tb]
\begin{center}
\unitlength 1mm
\begin{picture}(85,30)(0,5)
\put(0,0){\resizebox{85mm}{!}{\includegraphics{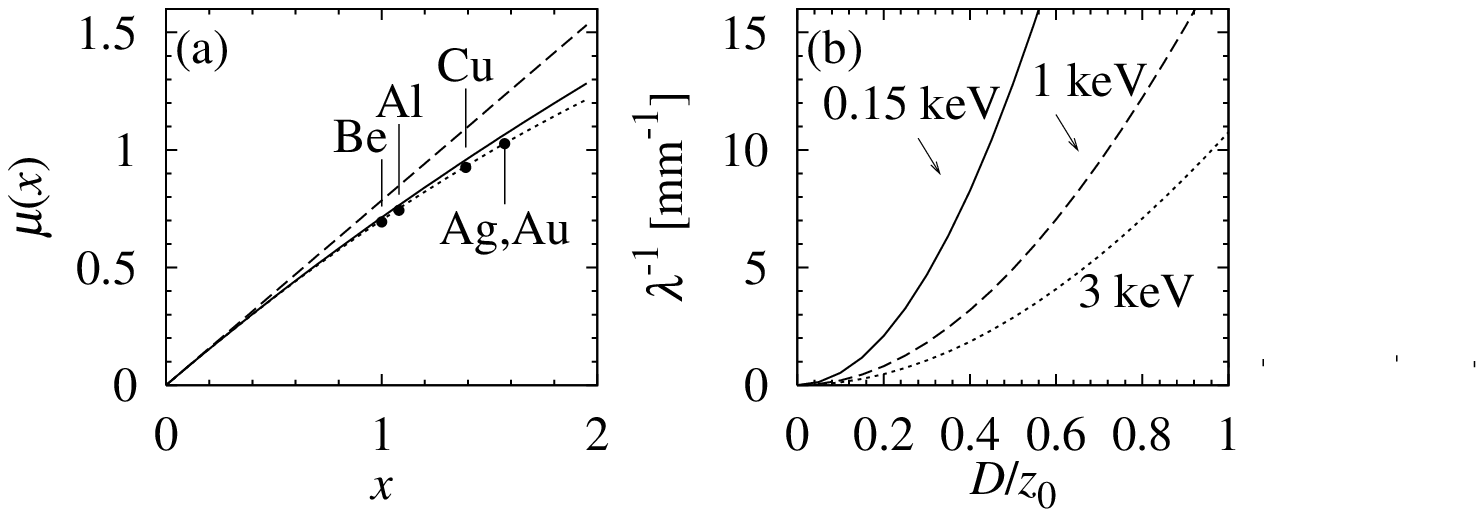}}}
\end{picture}
\end{center}
\caption{\label{fig:metal}(a) Solid line: the  ``material'' function 
$\mu(x)$ for metals with the values corresponding to a
few metals (assuming $\epsilon_{\mathrm{i}}=2$); dashed and dotted lines:
first and second order approximation according to
Eq.~(\ref{mu-expand}), respectively; points: the values for the
five metals corrected for the Coulomb and exchange correlations within the
Hubbard model.
(b) The inverse decoherence length as a
function of the $D/z_{0}$ ratio for gold at $T=293$ K, for three
values of the electron energy}
\end{figure}

\begin{figure}[tb]
\begin{center}
\unitlength 1mm
\begin{picture}(85,30)(0,5)
\put(0,0){\resizebox{85mm}{!}{\includegraphics{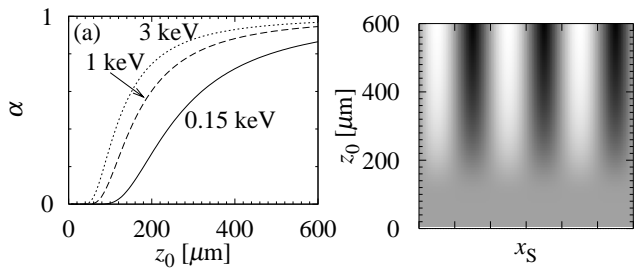}}}
\end{picture}
\end{center}
\caption{\label{fig:visib}(a) The visibility of interference fringes
as a function of the path distance from a gold plate of length $L=1$ cm,
assuming the separation between the
paths $D=10\;\mu$m, for three values of the electron energy. (b) A
simulation of interference fringes in an experiment with an
electron beam spread along $z$, in which the fringes for a range of
$z_{0}$ values are obtained simultaneously. Here  $D=10\;\mu$m,
$E=150$ eV, $x_{\mathrm{S}}$ is the coordinate along the screen.}
\end{figure}

The momentum integral in Eq.~(\ref{DRfinal}) may be written, using the
explicit expression (\ref{Ree}), in the form
\begin{displaymath}
\int dq_{z} 
\frac{\varepsilon_{2}(q_{z},\omega)}{\varepsilon_{1}^{2}(q_{z},0)}
=\omega\frac{2\pi\epsilon_{0}\hbar}{e^{2}}
\mu\left( 
\frac{me^{2}}{
2\pi\epsilon_{0}\epsilon_{\mathrm{i}}\hbar^{2}k_{\mathrm{F}}}
\right).
\end{displaymath}
The lower value of the integral over $q$ may be extended from
$\omega/v$ to 0. This is a negligible correction since
$\omega/v\ll k_{\mathrm{F}}$ and the expression under the integral is
regular at $q\to 0$.
The function 
\begin{eqnarray}\label{mu-m}
\lefteqn{\mu(x)=}\\
\nonumber 
&&\frac{x^{2}}{4}\int_{0}^{1} \frac{du}{u^{3}}
        \left[1+\frac{x}{4\pi u^{2}}\left(
            1+\frac{1-u^{2}}{2u}\ln\frac{1+u}{1-u}
        \right) \right]^{-2}
\end{eqnarray}
describes the properties of the metallic reservoir (the
material properties of the system). It is depicted in
Fig.~\ref{fig:metal}a.
Inserting the above result into Eq.~(\ref{DRfinal}) and using
Eq.~(\ref{gamma}) one gets the characteristic dephasing path length for a
metallic reservoir
\begin{equation}\label{metal}
\lambda^{-1}=-\frac{\Delta R}{L}
=\frac{k_{\mathrm{B}}T}{2\pi^{2}\hbar v}
\mu\left( 
\frac{me^{2}}{
2\pi\epsilon_{0}\epsilon_{\mathrm{i}}\hbar^{2}k_{\mathrm{F}}}
\right)\gamma\left( \frac{D}{z_{0}} \right).
\end{equation}
This result is shown for gold in Fig.~\ref{fig:metal}b for a few
electron energies. The dependence on material parameters is contained
in the function $\mu$ which, for many metals, differs
only slightly from that corresponding to gold (see
Fig.~\ref{fig:metal}a). For electrons with energy 150 eV the 
decoherence effect should be noticeable with a 1 cm plate
already for $D/z_{0}\sim 0.1$. In Fig.~\ref{fig:visib}, the visibility
of interference fringes for a specific system setup is shown, along
with a simulation of the fringes.

A standard way to quantitatively improve the description of the
screening response of an electron gas is to include correlations
between electrons leading to Coulomb and exchange hole around an
electron. This can be done within the Hubbard model,\cite{mahan00} where the
electron susceptibility (\ref{diel-expli}) is replaced by
\begin{eqnarray}\label{susc-H}
\chi_{\mathrm{el}}^{(\mathrm{H})}(q,\omega)=
\frac{\chi_{\mathrm{el}}(q,\omega)}{
1-G(q)\chi_{\mathrm{el}}(q,\omega)},
\end{eqnarray}
with \cite{mahan00}
\begin{displaymath}
G(q)=
\frac{1}{2}\frac{q^{2}}{q^{2}+k_{\mathrm{F}}^{2}+k_{\mathrm{TF}}^{2}},
\end{displaymath}
where $k_{\mathrm{TF}}$ is the Thomas--Fermi wave vector for a given
metal.
It turns out that this correction affects the value of
$\mu$ for typical metals only to a very little degree, as
shown in Fig.~\ref{fig:metal}(a) (points).

It may be interesting to obtain a closed formula for the decoherence
length $\lambda$ or for the decoherence time $\tau_{\mathrm{d}}$ 
and compare it to
the phenomenological prediction of Ref.~\onlinecite{anglin96}.
As can be seen in Fig. \ref{fig:metal}a, for most metals on has
\begin{displaymath}
\frac{me^{2}}{
2\pi\epsilon_{0}\epsilon_{\mathrm{i}}\hbar^{2}k_{\mathrm{F}}}\sim 1.
\end{displaymath}
In this range of
values the asymptotic expansion is valid
\begin{equation}\label{mu-expand}
\mu(x)\approx \frac{\pi}{4}x-\frac{1}{12}x^{2}
\end{equation}
(see Fig.~\ref{fig:metal}a).
Retaining only the linear term one finds 
\begin{displaymath}
\int dq_{z}
\frac{\varepsilon_{2}(q_{z},\omega)}{\varepsilon_{1}^{2}(q_{z},0)}
=\frac{\pi}{4}\frac{m}{\epsilon_{\mathrm{i}}\hbar k_{\mathrm{F}}}\omega,
\end{displaymath}
which leads to the approximate formula for the 
decoherence length $\lambda$,
\begin{equation}\label{lambda}
    \lambda^{-1}=
        \frac{k_{\mathrm{B}}T}{8\pi^{2}\hbar v}
        \frac{me^{2}}{\epsilon_{0}\epsilon_{\mathrm{i}}
            \hbar k_{\mathrm{F}}}
        \gamma\left(\frac{D}{z_{0}}\right).
\end{equation}
For $D\lesssim z_{0}$ we use the expansion from
Eq.~(\ref{gamma-expli}) and write for the decoherence time
$\tau_{\mathrm{d}}=\lambda/v$ 
\begin{equation}\label{tau-d}
\tau_{\mathrm{d}}^{-1}
=\frac{\pi}{32}\tau_{\mathrm{r}}^{-1}
\left( \frac{D}{\lambda_{\mathrm{dB}}} \right)^{2}, 
\end{equation}
where $\lambda_{\mathrm{dB}}=2\pi\hbar/\sqrt{2mk_{\mathrm{B}}T}$ is the
thermal de Broglie wave length of an electron and
\begin{equation}
\label{tau-r}
\tau_{\mathrm{r}}^{-1}=\frac{e^{2}}{2\pi\epsilon_{0}\epsilon_{\mathrm{i}}
	z_{0}^{2}\hbar k_{\mathrm{F}}}.
\end{equation}
The formula (\ref{tau-d}) is essentially of the same form as that given in
Ref. \onlinecite{anglin96}. However, here $\lambda_{\mathrm{dB}}$ refers to
the electrons in the metal and involves the mass of these electrons. 
In fact, decoherence results from the
perturbation of the reservoir state and not that of the system
\cite{zurek03}. Since the reservoir responds to the electromagnetic
field of the flying electron, its state may depend on its charge but
not on the mass. Therefore, it should be expected that particles of
different mass but equal charge will undergo the same decoherence.

The time constant $\tau_{\mathrm{r}}$ was originally interpreted as
the energy dissipation rate, i.e. the rate at which energy is
transferred from the electron to the reservoir. A similar
interpretation is possible here, within a semi-classical
approach. The electrons in the metal involved
in interactions with external fields are those near the Fermi
surface. The relative rate at which their energy changes is therefore
\begin{displaymath}
\tau_{\mathrm{r}}^{-1}=\frac{1}{E_{\mathrm{F}}}\frac{dE}{dt}.
\end{displaymath}
The variation of the energy of electrons is due to their acceleration by
the electric field ${\cal E}$ of the flying  electron the distance to which 
is roughly $z_{0}$, so that 
\begin{displaymath}
\frac{dE}{dt}=mv\dot{v}=mv\frac{e{\cal E}}{m}
\approx v_{\mathrm{F}}
\frac{e^{2}}{4\pi\epsilon_{0}\epsilon_{\mathrm{i}}z_{0}},
\end{displaymath}
where $v_{\mathrm{F}}$ is the electron speed at the Fermi surface and
the screening by ions has been included. Using the equalities 
$E_{\mathrm{F}}=\hbar^{2}k_{\mathrm{F}}^{2}/(2m)$ and 
$v_{\mathrm{F}}=\hbar k_{\mathrm{F}}/m$ one arrives at the formula
(\ref{tau-r}). 

In spite of the formal similarity of Eq.~(\ref{tau-d}) to the
phenomenological decoherence rate obtained from the energy losses due
to dc resistance and Joule heating \cite{anglin96}, the present theory
 actually describes a different mechanism of decoherence. The dc
resistance-based description assumes implicitly that formation of the
screening image charge is dissipationless (adiabatic), involving only
virtual transitions, and therefore reversible. The irreversible,
dissipative processes take place only due to carrier scattering as the
image charge moves beneath the metal surface. Since, at high
temperature, such scattering is mostly due to phonons, the which path 
information is effectively stored in lattice excitations. In contrast,
the quantum description presented here shows that already the process
of formation of the image charge is to a large extent dissipative,
even in the absence of carrier-phonon scattering. 
Since the reservoir response to the external charge involves only
excitations of low frequency but of arbitrary momenta the appropriate 
dielectric
function is different from the commonly used Drude limit. As a result,
there is no direct correlation between the decoherence effect and the
dc conductivity or resistivity. Instead, for metals the theory
predicts roughly inverse proportionality to the Fermi momentum which is the
only material parameter entering the result (apart from the electron
mass). 
Quantitative
comparison shows that for noble metals the rate given by
Eq.~(\ref{tau-r}) is many orders of magnitude higher than that resulting
from resistive dissipation. Therefore, the Ohmic resistivity effect is
of minor importance.

\section{Conclusions}
\label{sec:concl}

The decoherence effect on a single electron
travelling over a conducting plate was calculated in a model
reflecting the conditions of a currently performed experiment. 
The dissipative response of the 
electron gas in the plate, tending to screen the external charge, 
generates a trace in the plate. The resulting distinguishability of the
electron paths destroys the electron's ability to interfere. In order
to describe this effect quantitatively a fully quantum
model of the mutual interaction between the electron and the
conductive reservoir was formulated. 
The decoherence effect may be expressed by the
spectral density of the reservoir fluctuations which, in turn, is
related to its dissipative properties, i.e., to the imaginary part of
the inverse longitudinal dielectric function. The resulting
decoherence was described for a metallic reservoir but the qualitative
form of the result depends only on the universal linear
frequency dependence of its imaginary part. 
Therefore, the decoherence effect will be qualitatively
similar for any plate material (e.g., for semiconductors) as long as
the electron gas response is dominated by the low-frequency
sector. Whether this is the case for a specific system, will depend on
the interplay of the experimental conditions and material parameters.

The results presented in the paper explain the mechanism of electron
dephasing on a fully quantum level and give quantitative estimations
of the effect, depending on the material parameters and the geometry
of the experimental setup. Therefore, they should be helpful for
experimental studies in this fundamental field.  

\begin{acknowledgments}
The author is very grateful to Peter Sonnentag for comments and
discussions, and for supplying valuable information about the current
progress of the experiment. This work was
supported in parts by the Alexander von Humboldt Foundation 
and by the Polish Ministry of Scientific
Research and Information Technology under the (solicited) Grant
No. PBZ-MIN-008/P03/2003. 
\end{acknowledgments}

\appendix
\section{Explicit Markovian approximation}
\label{app:M}

In this Appendix, Eq.~(\ref{Rij2}) is re-derived in a way that clearly
displays the Markovian approximation by making an explicit use of the
short memory time $\tau_{\mathrm{mem}}$ of the screening response 
which leads to correlation
constants which are directly proportional to time. Now we consider the
evolution during the time segment $t_{0}<t<t_{0}+\Delta t$ during
which the electron is moving over the plate. We assume that $\Delta t$
is much longer than $\tau_{\mathrm{mem}}$ but short enough for the
accumulated perturbation to be small. Since 
$\tau_{\mathrm{mem}}\sim \hbar/E_{\mathrm{F}}$ is of order of
femtoseconds and the time of flight over a 1~cm plate is in the
nanosecond range such time-slicing is always possible in the
experimentally interesting situation of non-complete dephasing.

Let us start again by substituting Eqs.~(\ref{psi-r-r}) and (\ref{dens-cor}) into
Eq.~(\ref{Rij}). Now, however, we first integrate over $y_{1,2}'$,
which yields $4\pi^{2}\delta(p_{y}+q_{y})\delta(p_{y}'-q_{y})$, and over 
$x_{1,2}'$, which yields 
$4\pi^{2}\delta^{(1)}_{L}(p_{x}+q_{x})\delta^{(1)}_{L}(p_{x}'-q_{x})$,
where $\delta_{L}^{(1)}(q)=\sin(qL/2)/(\pi q)$. Based on the same
argument as before, for $L\gg
z_{0}$ the latter may me replaced with a Dirac delta. The trivial
integration over $p_{x,y}$ and $p_{x,y}'$ yields a formula
containing the time integration in the form
\begin{eqnarray*}
\lefteqn{\int_{t_{0}}^{t}d\tau\int_{t_{0}}^{t}d\tau'
e^{-iq_{x}v(\tau-\tau')}\mathcal{M}_{\bm{q}}(\tau-\tau')}\\
&\approx & \int_{t_{0}}^{t}d\tau\int_{-\infty}^{\infty}du
e^{-iq_{x}vu}\mathcal{M}_{\bm{q}}(u)\\
&=&-2\pi\Delta t\coth\frac{\hbar\beta\omega}{2} 
\im \varepsilon^{-1}(\bm{q},-q_{x}v),
\end{eqnarray*}
where we defined the memory function
\begin{displaymath}
\mathcal{M}_{\bm{q}}(t)=\int_{0}^{\infty}d\omega e^{-i\omega t}
\coth\frac{\hbar\beta\omega}{2} \im \varepsilon^{-1}(\bm{q},\omega)
\end{displaymath}
and extended the limits of the second integration because 
$\mathcal{M}_{\bm{q}}(t)$ decays on time scales much shorter than 
$\Delta t$.

Using this result one obtains 
\begin{eqnarray*}
\lefteqn{ R_{ij} =}\\
&& \frac{\Delta t e^{2}}{2\pi^{2}\epsilon_{0}\hbar}
    \frac{1}{V}\sum_{q_{x}<0,q_{y},q_{z}}q^{2}
\coth \frac{\hbar\beta |q_{x}| v}{2}
    \im\left[-\frac{1}{\varepsilon(\bm{q},|q_{x}|v)}\right] \\
&&\times    
  e^{iq_{y}(y_{i}-y_{j})-\frac{1}{2}\left[
    (l_{x}q_{x})^{2}+(l_{y}q_{y})^{2} \right]}
    |I_{z}|_{\omega=|q_{x}|v}^{2},
\end{eqnarray*}
which is identical to the $L\to\infty$ limit of Eq.~(\ref{Rij2}) with
$L=v\Delta t$.

%\bibliographystyle{prsty}
%\bibliography{abbr,quantum}

\end{document}